\documentclass[12pt,preprint]{aastex}
\shorttitle{Nereid's Phase Function}
\shortauthors{Schaefer \& Tourtellotte}

\begin{document}

\title{The Shape of Nereid's Phase Function Changed From 1998 to 1999 and 2000}

\author{Bradley E. Schaefer\altaffilmark{1}}
\affil{Department of Physics, Yale University,
    New Haven CT 06520-8121}

\and

\author{Suzanne W. Tourtellotte}
\affil{Department of Astronomy, Yale University, New Haven, CT 06520-8121}

\altaffiltext{1}{Current address is Department of Astronomy, C-1400, 
University of Texas, Austin TX 78712}

\begin{abstract}
We present well-sampled light curves of Nereid for 1999 and 2000 which show symmetric shapes centered on opposition as is characteristic of a large opposition surge.  Surprisingly, the phase functions for 1999 and 2000 are significantly different from the well-measured phase function for the year 1998, with the 1999 and 2000 curves being more peaked at low phase angles and substantially brighter at high phase angles.  We know of neither precedent nor explanation for this mystery on Nereid.
\end{abstract}



\section{Introduction}

	Nereid has an unusual orbit around Neptune ($e=0.75$, $P=360$ 
days, and $i=28 \degr$) which suggests that it might be a captured 
Kuiper Belt Object \citep{scs95,scs00}.  Nereid also 
has the most unusual known photometric history of all objects in 
the Solar System \citep{scs88,scs00,sct01}.  From 1987 to 1991, 
several groups found that 
Nereid had fast photometric variations with amplitudes up to around 
one magnitude \citep{scs88,scs00,wjt91,bul89}.  From 1991 to 1997, 
Nereid's 
amplitude was smaller at $\sim 0.4$ mag \citep{scs00},
with highly significant evidence that it showed fast variations 
\citep{bgm97,brw98}.  In 1998, we observed 
Nereid on 52 nights and found that Nereid displayed no significant 
variability other than a very large opposition surge \citep{sct01}.

	This unique behavior of large-and-small variations that 
change from year to year could be caused by chaotic rotation \citep{dob95}, 
much like Hyperion \citep{kla89}.  This 
possibility is appealing because Nereid has the necessary ingredients 
of a $>1\%$ out-of-round shape (due to its small size) and a highly 
eccentric orbit around a planet with a large J2 component in its 
gravity.  Unfortunately, Dobrovolskis has shown that Nereid's rotation 
will be chaotic only if its rotational period is longer than roughly 
two weeks, and this requirement is not consistent with the observed 
brightness changes on time scales of one day.  Thus, the cause of 
Nereid's variations is not known \citep{sct01}.

	The 1998 light curve \citep[see Figure 1]{sct01} displayed 
no significant brightness changes other than those associated with 
the normal changing of the solar phase angle, $\alpha$, and the 
opposition surge.  The phase function (see Figure 2) shows an 
opposition surge of 0.52 magnitudes over a range of 2$\degr$ in 
phase and this is among the largest known in our Solar System.  
The shape of the phase function is definitely not linear, while 
a broken line fits well.

	Given Nereid's long history of changes in its variability, 
it is prudent to continue photometric monitoring.  The Yale 1-m 
telescope on Cerro Tololo is operated in a queue mode by a resident 
operator, and hence is perfect for the long-term synoptic study of 
Nereid.  So again we have used this telescope to monitor Nereid, 
this time during its 1999 and 2000 oppositions.  We find that 
Nereid's phase function has significantly changed its shape from 
1998 to 1999 and 2000.

\section{Photometry}

	All photometry was taken with the Yale 1-m telescope 
(operated by the YALO consortium) with the ANDICAM CCD camera 
which has a $2048 \times 2048$ array of $0.29 \arcsec$ pixels.  
All exposures were through a standard V-band filter for 900 
seconds of exposure.  A complete description of our analysis 
procedures are presented in \citet{sct01}.

	In 1999, we obtained a total of 54 usable images on 54 
separate nights in a 130 day interval between 20 June 1999 (JD 
2451350) and 28 October 1999 (JD 2451480).  Nereid passed through 
opposition on 27 July 1999 (JD 2451386) and was at maximum phase 
angle on 26 April 1999 (JD 2451294) and 25 October 1999 (JD 2451476).

	In 2000, we obtained a total of 14 usable images on 14 
separate nights in a 102 day interval between 22 July 2000 (JD 
2451748) and 1 November 2000 (JD 2451850).  Nereid passed through 
opposition on 28 July 2000 (JD 2451753) and was at maximum phase 
angle on 28 April 2000 (JD 2451662) and 26 October 2000 (JD 2451843).

	All our V-band photometry is presented in Table 1.  The 
first column gives the heliocentric Julian Date of the start of 
the exposure.  The second column gives solar phase angle, $\alpha$, 
in degrees as given by the JPL Horizons program 
(http://ssd.jpl.nasa.gov/cgi-bin/eph).  The third column gives our 
observed V magnitude and one-sigma uncertainty.  The fourth column 
gives the V magnitude corrected to the 1998 opposition distance of 
29.123 AU as $V_{OPP}=V-5 \log (\Delta /29.123)$ where $\Delta$ is 
the Earth-Nereid distance in units of AU, to allow direct comparisons 
with the 1998 phase function.  The error bars for $V_{OPP}$ are the 
same as for $V$ in the previous column.  (The total range of variation 
of the Sun-Nereid distance over the observed time intervals in 1998-2000 
was only 0.08 AU, thus resulting in a negligible correction of under 
0.003 mag from the mean.)  The light curves are displayed in Figure 1.

	We notice that the one measurement with $\alpha < 0.11 \degr$ 
in 1999 (on JD 2451386.829) is bright by $\sim$ 0.2 mag when compared 
to observations taken a few days before and afterwards.    We have 
carefully examined this image for bad columns, measured Nereid's image 
shape for evidence of cosmic rays, checked other deeper images for 
background stars, and checked Nereid's position on the CCD chip with 
other images for hot pixels, all with the result that we have no 
grounds for impeaching this 'bright' magnitude.  (Actually, similar 
checks were performed on {\it all} our Nereid images.)  The coincidence of 
this bright Nereid with phase angle zero suggests that this might be 
evidence of a narrow spike in Nereid's phase function (much like that 
for four of the Uranian moons \citep{bgm92}.  Yet 
such a spike does not appear in the well-sampled 1998 light curve, so 
the 'spike' would have to be time variable if it exists.

\section{Phase Function}

	In 1999, the observed light curve comes to a peak around 26 
July 1999 and a minimum in October 1999 with a symmetric light curve 
around the peak.  In 2000, the observed light curve comes to a peak 
around 26 July 2000 and a minimum in October 2000.  The coincidence 
of the observed maximum and minimum dates with the opposition and 
quadrature dates as well as the symmetric light curves are the 
hallmarks of brightness variations caused by the opposition surge.  
Thus, it appears that (as in 1998), most of the variations arises 
from simple phase effects.  The phase functions for 1999 and 2000 
are presented in Figure 2.                     

	We first tried to fit the 1999 and 2000 data to linear and 
broken-line functional forms.  For the 1999 data, the broken-line 
form is strongly preferred over the linear form.  The best fit is 
$V_{OPP}=18.97+1.09 \alpha$ for $\alpha < 0.29 \degr$ and 
$V_{OPP}=19.25+0.13 \alpha$ for $\alpha > 0.29 \degr$ with a 
chi-square of 40.3 for 50 degrees of freedom.  The 1999 data is 
strongly inconsistent with the 1998 broken-line fit \citep{sct01}.  
The 1999 and 2000 data have nearly identical fits.  

	Fits to lines and broken-lines have the virtue of simplicity, 
but they have no physical motivation.  \citet{hap93} developed a 
physical model that predicts the opposition surge component of the 
phase function to vary as 
$V_{OPP}=V_0 - 2.5 \log \{ [1+B_0/(1+0.5 \alpha /h)]/[1+B_0] \} $ in 
the small angle limit.  Here, $V_0$ is the V-band magnitude at zero 
phase angle, $B_0$ is a measure of the amplitude of the opposition 
surge, and $h$ is the angular width of the opposition surge.  The use 
of this physical model has the disadvantage that it might not be a 
complete description of the physics (see discussion in the next 
section).  Also, the physical model does not describe the observed 
variations over small ranges of phase angles, in particular at low 
phase angles.

	The best fit for the 1999 data has $V_0 = 19.01$ mag, 
$B_0 = 0.72$, and $h=0.19\degr$ for a chi-square of 48.8 with 51 
degrees of freedom.  The best fit for the 2000 data has 
$V_0 = 19.07$ mag, $B_0 = 0.78$, and $h=0.34 \degr$ for a chi-square 
of 9.6 with 11 degrees of freedom, although the 2000 errors are 
relatively large and correlated.  The fit parameters are similar for 
1999 and 2000.  The best way to show this is by evaluating the 
chi-squares for each year with the other year's fit.  The 1999 data 
have a chi-square of 52.1 for the 2000 best fit; while the 2000 data 
have a chi-square of 16.4 for the 1999 best fit.  The reduced 
chi-square values are near unity and F-Tests show that the cross-fits 
are not significantly different from each other.  A joint fit of the 
1999+2000 data reveal a best fit with $V_0 = 19.04$ mag, $B_0 = 0.73$, 
and $h=0.25 \degr$ for a chi-square of $49.4+12.0=61.4$ with 65 degrees 
of freedom.  Again, the differences in chi-square are not significant.  
Hence, it appears that the phase functions in 1999 and 2000 were similar 
if not identical.  So we will adopt the best fit to the Hapke model 
for the joint 1999+2000 data as a good description of the phase 
function for both years.

	Is the 1998 phase function \citep{sct01} the 
same as what we are here reporting for 1999 and 2000?  The 1998 data 
were fitted to the Hapke model with  $V_0 = 19.07$ mag, $B_0 = 1.32$, 
and $h=0.52 \degr$ for a chi-square of 70.0 with 54 degrees of freedom.  
Both the 1998 and the 1999+2000 model phase functions are drawn in all 
three panels of Figure 2.  In examining this figure, the answer is 
obvious that the observed phase function has changed from 1998 to 
1999+2000.  In particular, the observed 1999+2000 phase function appears 
to be  more peaked at low phases and substantially brighter at large 
phases when compared to the observed 1998 phase function.  

	The 1998 phase function and light curve is flat at low phase 
angles, while the 1999+2000 phase function and light curve is strongly 
peaked to opposition.  To be quantitative, in bins of 0.05$\degr$ phase 
from 0.00$\degr$ to 0.30$\degr$, the 1998 observed values are 
$19.143  \pm 0.026$, $19.137 \pm 0.021$, $19.105 \pm 0.020$, 
$19.14 \pm 0.05$, $19.118 \pm 0.027$, and $19.199 \pm 0.028$ mag.  
Similarly the 1999+2000 binned phase function is $19.064 \pm 0.022$ 
(or $19.087 \pm 0.024$ if the point nearest opposition is arbitrarily 
ignored), $19.12 \pm 0.03$, $19.155 \pm 0.019$, $19.180 \pm 0.018$, 
$19.173 \pm 0.026$, and $19.253 \pm 0.032$ mag.  All 1998 points with 
$0.00 \degr < \alpha < 0.26 \degr$ are consistent with a perfectly 
flat phase curve at $19.126 \pm 0.011$ mag.  The 1999+2000 magnitudes 
form a simple linear decline that is not consistent with a flat phase 
function.  Formal fits for the slopes yield $0.12 \pm 0.12$ mag deg$^{-1}$ 
for 1998 and $0.73 \pm 0.14$ mag deg$^{-1}$ for 1999+2000 ($0.63 \pm 0.14$ 
mag deg$^{-1}$ if the near opposition datum is excluded despite the lack 
of grounds).  

	At large phases with $\alpha > 0.85 \degr$, all 1998 points lie 
{\it below} the 1999+2000 fit and all 1999+2000 points lie {\it above} 
the 1998 fit. (Both statements have a single exception which in both 
cases is the one point with the largest error bar.)  Taking points with 
$\alpha > 1.5 \degr$, the weighted averages for 1998 and 1999+2000 are 
$19.551 \pm 0.013$ and $19.485 \pm 0.010$ mag respectively, for a 
difference of $0.066 \pm 0.016$ mag which is significant at the 4.0 
sigma level.  Hence, the observed phase function at large phases changed 
significantly after 1998.

	Overall, the change in the shape of the phase function can be 
tested by comparing the fits to the Hapke model.  The 1998 phase function 
when applied to the 1999+2000 data yields a chi-square of $119.3+19.9=139.2$ 
for 68 degrees of freedom (versus a chi-square of 61.4 for the 1999+2000 
phase function), and this strongly rejects the 1998 function for the 
1999+2000 data.  Similarly, the 1999+2000 phase function when applied to 
the 1998 data yields a chi-square of 120.7 for 57 degrees of freedom 
(versus a chi-square of 70.0 for the 1998 phase function), and this 
strongly rejects the 1999+2000 phase function for the 1998 data.  Thus, 
Nereid's phase function changed from 1998 to 1999+2000.

	This is an unprecedented result, so it is prudent to consider 
whether there is any realistic possibility of observational error.  For 
example, what if we adopted magnitudes for the 1998 comparison stars used 
at high phase angles that are too faint by 0.07 mag due to some calibration 
error?  (This would still not resolve the difference in the slope at low 
phases, so a separate error would have to be postulated.)  A 0.07 mag error 
is far larger than any reasonable statistical error, and our repeated 
independent derivation of all the calibrations argues against numerical 
errors.  The same telescope, CCD chip, filters, analysis procedures, and 
many of the same Landolt standard stars were used throughout each of the 
three years.  The presence of undetected light cirrus clouds that might 
dim the comparison star images in comparison to the standard star images 
is ruled out since the $\alpha > 1.3 \degr$ comparison star data in 1998 
was independently calibrated against Landolt standard stars on three 
separate nights and these all agree to within $\sim$ 0.015 mag.  
Similarly, the comparison stars for the 1999 data with $\alpha > 1.3 \degr$ 
were independently calibrated on three nights with the YALO telescope and 
one night with the McDonald 2.1-m telescope.  Nereid crossed its own path 
on nights around 20 June 1998 and 20 October 1999, for which our independent 
calibrations of the comparison stars are in agreement to within 
$0.03 \pm 0.03$ mag.  In all, we have strong reasons to be confident in 
our photometry and thus we conclude that Nereid's phase function did 
indeed change from 1998 to 1999+2000.

	Nereid has a long history of photometric variations, so could the 
phase function changes actually be small amplitude long-term changes?  
Conceptually, it is difficult to know how the two cases can be 
distinguished, especially when we do not know the cause of the 
variability.  Certainly we could postulate changes in Nereid's 
overall brightness (but not its surface texture) that happen to 
make it faint in September and October 1998 (or bright in September 
and October of both 1999 and 2000) as well as fade Nereid by $\sim$ 0.2 
mag in the week centered on the 1998 opposition (or brightened Nereid in 
the week centered on both the 1999 and 2000 oppositions).  But such a 
postulate would require two rather improbable coincidences.  The first 
coincidence is that the fading in 1998 is centered on the opposition 
date, since this time is only special from an Earthly viewpoint.  (At 
this time, Nereid is far from Neptune and its 360 day orbital period 
means that it will be at roughly the same orbital phase for the 
oppositions from 1998-2000.)  The second coincidence is that Nereid's 
brightness always returns to the same magnitude the five times when 
$0.4 \degr < \alpha < 0.8 \degr$ (early July 1998, middle August 1998, 
early July 1999, middle August 1999, and middle August 2000).  The point 
to these coincidences is that Nereid's brightness appears to be correlated 
with the position of the Earth.  This correlation only makes sense if the 
observed variations are caused by a changing phase function which is 
explicitly a function of the Earth's position.  So again, we are 
concluding that Nereid's phase function changed from 1998 to 1999+2000.

\section{Implications}

The best fit for the 1998 data has $B_0$ significantly larger than 
unity, whereas the model of Hapke requires $B_0 \leq 1$.  This is 
actually a fairly broad problem, as many objects throughout the 
Solar System have this same difficulty \citep{hvh97}.  
The likely solution is that the true situation is more complex than 
given in the basic model, perhaps involving a combination of 
coherent-backscattering plus strongly backscattering single-particle 
phase functions.

The narrow width of the opposition surge for Nereid has precedents from 
four Uranian moons (Miranda, Ariel, Titania, and Oberon) with high 
albedos \citep{bgm92} as well as for Europa \citep{dom91,hel98}.  
In general, simple 
shadow-hiding cannot account for such narrow widths without presuming 
a very low filling factor for objects in the regolith, with $\sim 1 \%$ 
being a typical value (cf. Hapke 1993, equations 8.84 and 8.85).  Such a 
low value seems implausible, although \citet{hap83} has proposed that 
scatterers might be concentrated towards the surface of a relatively 
deep and transparent medium.  Nevertheless, the narrow component in 
the opposition surge is generally thought to arise from 
coherent-backscattering \citep{hap93}.  In this case, 
$h \approx \lambda / (4 \pi L)$ where $\lambda$ is the wavelength of the 
light and $L$ is the photon mean free path in the medium (Hapke 1993, 
equation 8.88).  For visible light ($\lambda=0.55~\micron$) and 
$h=0.0043$ rad ($0.25 \degr$), $L=10.2~\micron$.

We interpret the changes displayed in Figure 2 as due to a variable 
phase function, but other explanations are possible.  It is always 
possible to postulate subtle observational errors that can explain 
systematic offsets of $\sim$ 0.1 mag.  Nevertheless, in our case, 
the stability of our equipment and procedures along with our many 
independent calibrations and cross checks gives us strong confidence 
that our results are reliable.  It is always possible to postulate some 
arbitrary fading and brightening caused by whatever mechanism has 
previously made Nereid a variable, such that the light curves in Figure 1 
are produced without changing the surface texture.  Nevertheless, such 
postulated changes appear to be correlated with the Earth's position due 
to the low-phase differences being centered on opposition and due to the 
return to the same magnitude on five occasions when the phase is 
$0.4 \degr < \alpha < 0.8 \degr$.  This connection with the Earth argues 
that the light curve changes arise from phase function changes.  These 
arguments give us the grounds to identify the light curve changes as 
arising from changes in the phase function of Nereid.

A central mystery of Nereid remains to explain its large-and-small 
variations, only now there is the associated mystery of how it can 
change its phase function on a time scale of one year.  Presumably, 
as the phase function depends on the texture of the surface, then 
Nereid's texture must have changed.  Could this have been caused by 
the deposition of particles or the formation of ice crystals on the 
surface?  If so, then what could have caused this surface change?  
How is this resurfacing mechanism related to the brightness changes 
seen by so many groups?  Or could chaotic rotation of Nereid have 
moved a different hemisphere into view, with each hemisphere having 
different textures?  Unfortunately, we do not have answers to any of 
these questions.

We know of no precedents for significant changes in the shape of the 
phase function of Solar System objects.  But this statement is not 
as strong as might be expected, since few objects in our Solar System 
have been examined with well-sampled phase functions over more than 
one opposition with measures that can be directly compared.  Likely, 
Nereid's variable phase function is related somehow to its previously 
reported variations, but the causes of both types of changes are 
unknown.  To solve this mystery, a series of well-sampled light 
curves is certainly needed, ideally in several different bands and 
accompanied by spectroscopy.

\clearpage

\begin{deluxetable}{cccc}
\tabletypesize{\scriptsize}
\tablecaption{Brightness of Nereid in 1999 and 2000. \label{tbl-1}}
\tablewidth{0pt}
\tablehead{
\colhead{Julian Date} & \colhead{$\alpha$}   & \colhead{$V$}   &  \colhead{$V_{OPP}$}
}
\startdata

2451350.914	&	1.094	&	19.42	$ \pm $	0.06	&	19.41	\\
2451352.909	&	1.039	&	19.44	$ \pm $	0.05	&	19.43	 \\
2451354.916	&	0.984	&	19.36	$ \pm $	0.03	&	19.35	 \\
2451355.902	&	0.955	&	19.39	$ \pm $	0.03	&	19.38	 \\
2451365.891	&	0.657	&	19.26	$ \pm $	0.04	&	19.26	 \\
2451366.881	&	0.628	&	19.30	$ \pm $	0.04	&	19.30	 \\
2451367.877	&	0.596	&	19.25	$ \pm $	0.05	&	19.25	 \\
2451379.843	&	0.210	&	19.16	$ \pm $	0.03	&	19.16	 \\
2451380.844	&	0.177	&	19.14	$ \pm $	0.03	&	19.14	 \\
2451381.847	&	0.144	&	19.16	$ \pm $	0.03	&	19.16	 \\
2451382.846	&	0.111	&	19.11	$ \pm $	0.03	&	19.11	 \\
2451386.829	&	0.024	&	18.95	$ \pm $	0.05	&	18.95	 \\
2451389.847	&	0.122	&	19.23	$ \pm $	0.06	&	19.23	 \\
2451390.839	&	0.154	&	19.18	$ \pm $	0.08	&	19.18	 \\
2451391.869	&	0.189	&	19.28	$ \pm $	0.06	&	19.28	 \\
2451392.840	&	0.220	&	19.19	$ \pm $	0.04	&	19.19	 \\
2451393.837	&	0.253	&	19.20	$ \pm $	0.05	&	19.20	 \\
2451394.830	&	0.286	&	19.29	$ \pm $	0.04	&	19.29	 \\
2451396.842	&	0.351	&	19.31	$ \pm $	0.05	&	19.31	 \\
2451399.811	&	0.448	&	19.37	$ \pm $	0.04	&	19.37	 \\
2451401.817	&	0.512	&	19.35	$ \pm $	0.03	&	19.35	 \\
2451402.807	&	0.543	&	19.31	$ \pm $	0.03	&	19.31	 \\
2451403.805	&	0.574	&	19.30	$ \pm $	0.04	&	19.30	 \\
2451406.809	&	0.668	&	19.38	$ \pm $	0.06	&	19.38	 \\
2451407.793	&	0.699	&	19.36	$ \pm $	0.03	&	19.36	 \\
2451408.800	&	0.730	&	19.37	$ \pm $	0.04	&	19.37	 \\
2451410.785	&	0.790	&	19.39	$ \pm $	0.04	&	19.39	 \\
2451412.787	&	0.850	&	19.37	$ \pm $	0.06	&	19.36	 \\
2451432.738	&	1.376	&	19.40	$ \pm $	0.04	&	19.38	 \\
2451433.709	&	1.398	&	19.42	$ \pm $	0.04	&	19.40	 \\
2451441.730	&	1.562	&	19.48	$ \pm $	0.09	&	19.45	 \\
2451443.682	&	1.597	&	19.58	$ \pm $	0.15	&	
19.55	 \\
2451444.670	&	1.615	&	19.34	$ \pm $	0.09	&	19.31	 \\
2451445.670	&	1.631	&	19.52	$ \pm $	0.07	&	19.49	 \\
2451446.674	&	1.648	&	19.45	$ \pm $	0.10	&	
19.42	 \\
2451449.660	&	1.694	&	19.50	$ \pm $	0.05	&	19.46	 \\
2451451.657	&	1.723	&	19.54	$ \pm $	0.08	&	
19.50	 \\
2451454.652	&	1.761	&	19.47	$ \pm $	0.04	&	
19.43	 \\
2451455.646	&	1.773	&	19.56	$ \pm $	0.05	&	
19.52	 \\
2451458.634	&	1.804	&	19.53	$ \pm $	0.04	&	
19.48	 \\
2451460.662	&	1.824	&	19.59	$ \pm $	0.05	&	
19.54	 \\
2451461.674	&	1.832	&	19.53	$ \pm $	0.04	&	
19.48	 \\
2451463.589	&	1.847	&	19.53	$ \pm $	0.04	&	
19.48	 \\
2451465.616	&	1.861	&	19.55	$ \pm $	0.04	&	
19.49	 \\
2451466.679	&	1.867	&	19.53	$ \pm $	0.05	&	
19.47	 \\
2451467.615	&	1.872	&	19.54	$ \pm $	0.05	&	
19.48	 \\
2451468.669	&	1.877	&	19.55	$ \pm $	0.10	&	
19.49	 \\
2451471.581	&	1.887	&	19.54	$ \pm $	0.06	&	19.48	 \\
2451472.576	&	1.889	&	19.54	$ \pm $	0.07	&	
19.48	 \\
2451475.577	&	1.893	&	19.49	$ \pm $	0.06	&	
19.42	 \\
2451477.572	&	1.893	&	19.58	$ \pm $	0.05	&	
19.51	 \\
2451478.590	&	1.892	&	19.56	$ \pm $	0.05	&	19.49	 \\
2451479.576	&	1.890	&	19.57	$ \pm $	0.04	&	
19.50	 \\
2451480.580	&	1.888	&	19.60	$ \pm $	0.04	&	19.53	 \\
2451748.843	&	0.162	&	19.18	$ \pm $	0.04	&	19.18	 \\
2451750.839	&	0.096	&	19.12	$ \pm $	0.03	&	19.12	 \\
2451752.841	&	0.031	&	19.07	$ \pm $	0.03	&	19.07	 \\
2451754.827	&	0.037	&	19.10	$ \pm $	0.03	&	19.10	 \\
2451756.847	&	0.104	&	19.17	$ \pm $	0.04	&	19.17	 \\
2451758.761	&	0.166	&	19.18	$ \pm $	0.03	&	19.18	 \\
2451766.801	&	0.429	&	19.24	$ \pm $	0.04	&	19.24	 \\
2451768.805	&	0.493	&	19.36	$ \pm $	0.06	&	19.36	 \\
2451782.757	&	0.921	&	19.38	$ \pm $	0.04	&	19.37	 \\
2451784.749	&	0.979	&	19.35	$ \pm $	0.04	&	19.34	 \\
2451808.691	&	1.552	&	19.45	$ \pm $	0.05	&	19.42	 \\
2451810.684	&	1.588	&	19.52	$ \pm $	0.03	&	19.49	 \\
2451819.599	&	1.728	&	19.59	$ \pm $	0.05	&	19.55	 \\
2451850.567	&	1.880	&	19.55	$ \pm $	0.04	&	19.47	 \\

\enddata

\end{deluxetable}

\clearpage 

\begin{figure}
\columnwidth=0.5\columnwidth
\plotone{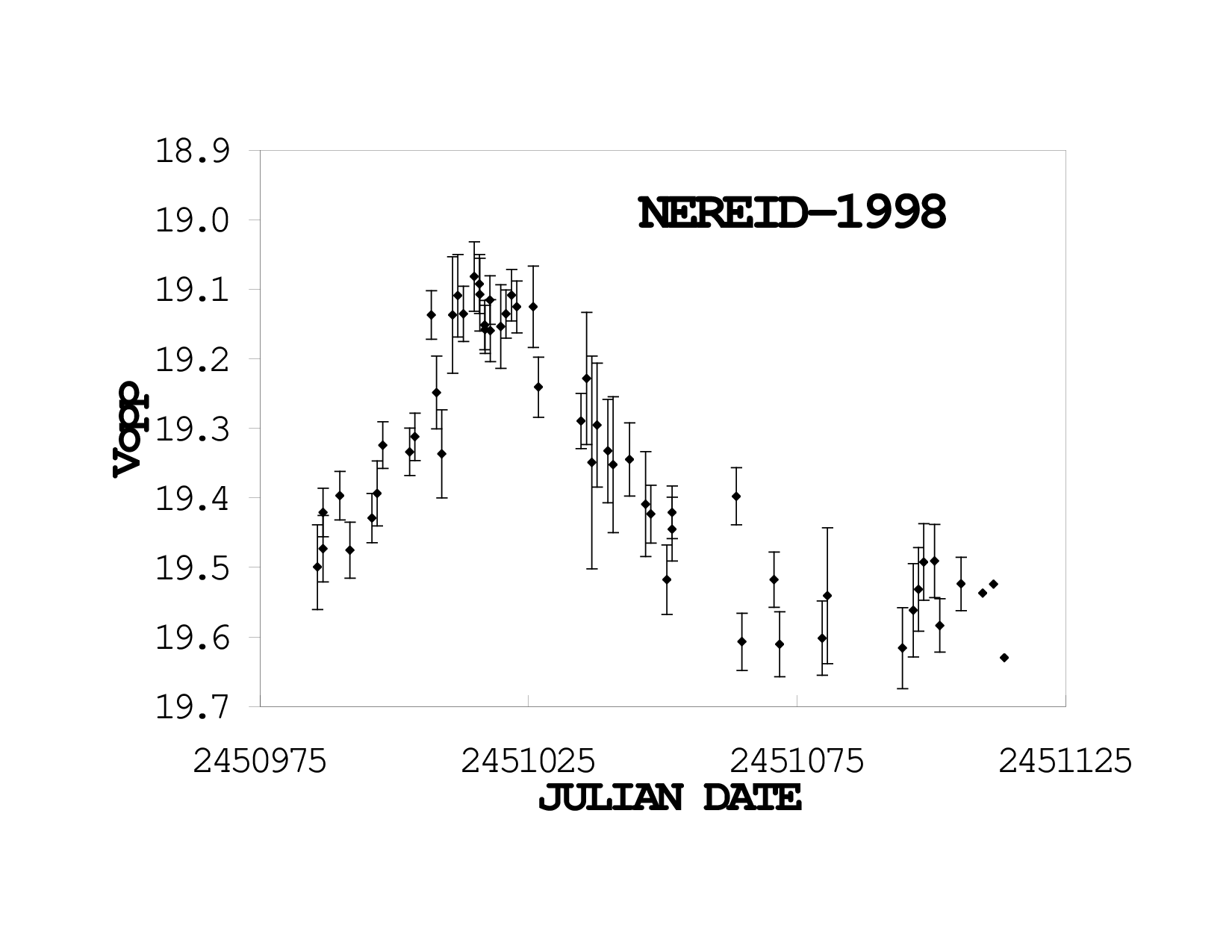}
\vspace{-0.3in}
\plotone{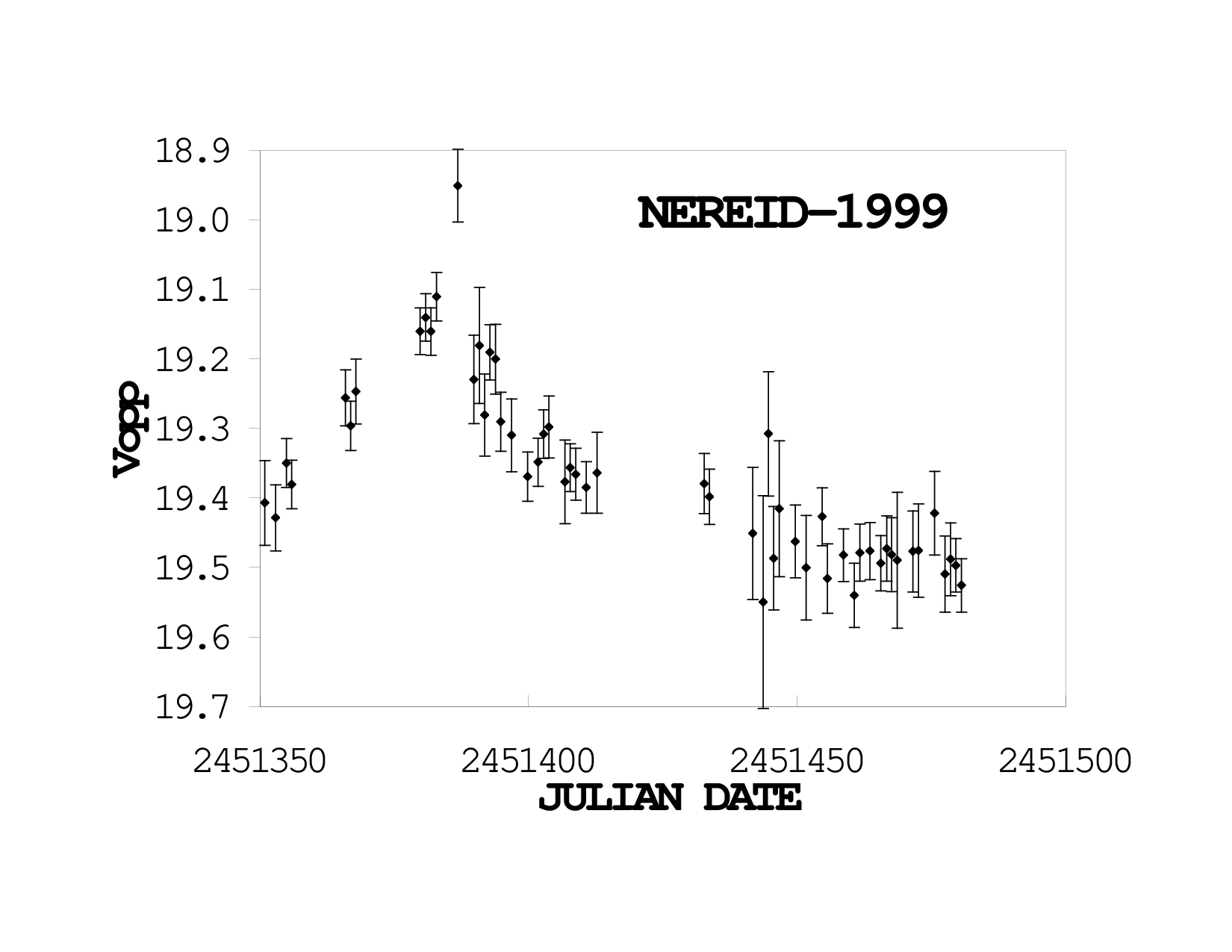}
\vspace{-0.3in}
\plotone{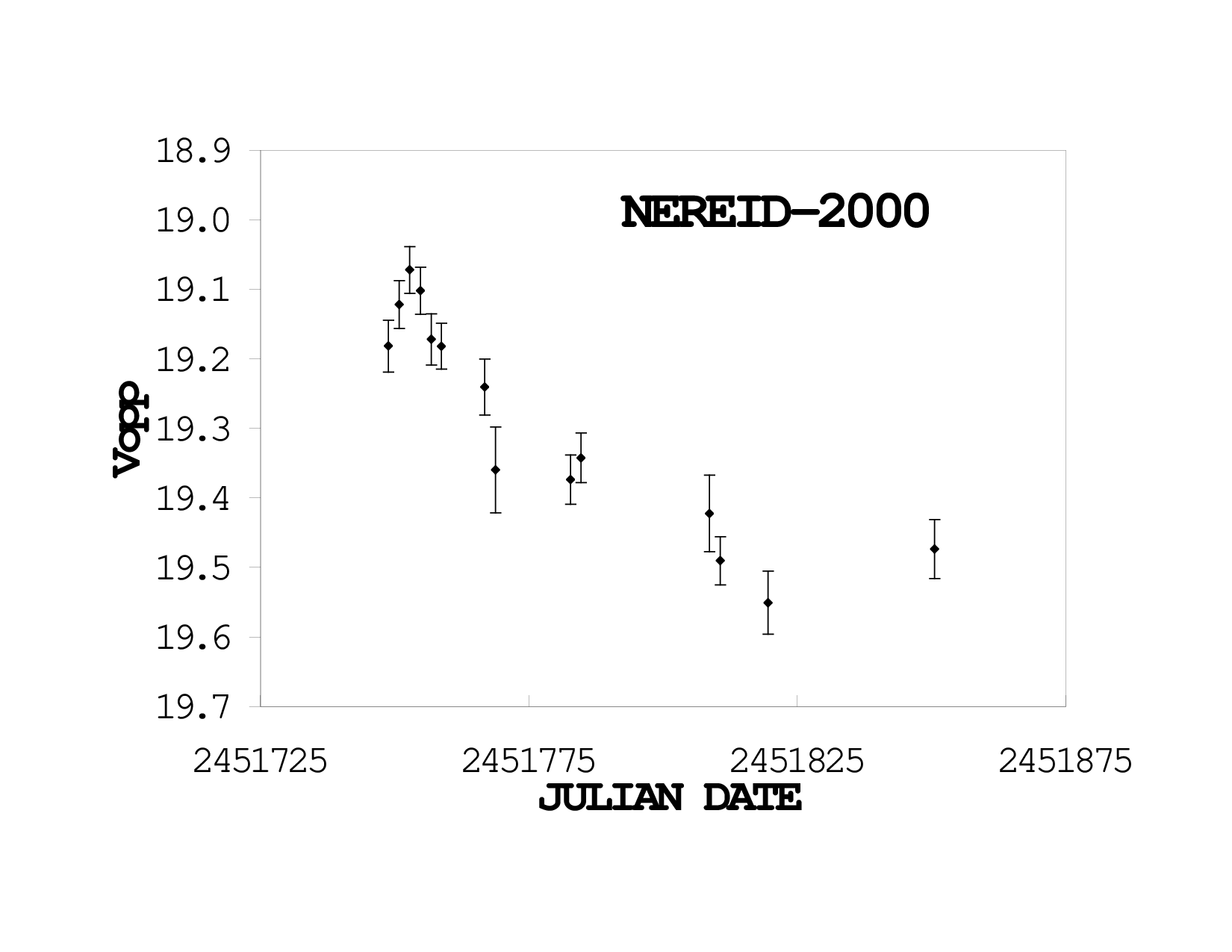}
\vspace{-0.1in}
\columnwidth=2\columnwidth
\caption{
Nereid light curves in 1998, 1999, and 2000.
The light curves for the three years show Nereid to have a smooth 
brightening to the date of opposition and then a smooth decline until 
a minimum around the time of quadrature.  This is the hallmark of the 
opposition surge, for which Nereid has a very large and narrow surge 
when compared to most other bodies in the Solar System.  Note that the 
light curve for 1998 is flat at opposition while the light curves for 
1999 and 2000 are sharply peaked.
\label{fig1}}
\end{figure}

\clearpage

\begin{figure}
\columnwidth=0.5\columnwidth
\plotone{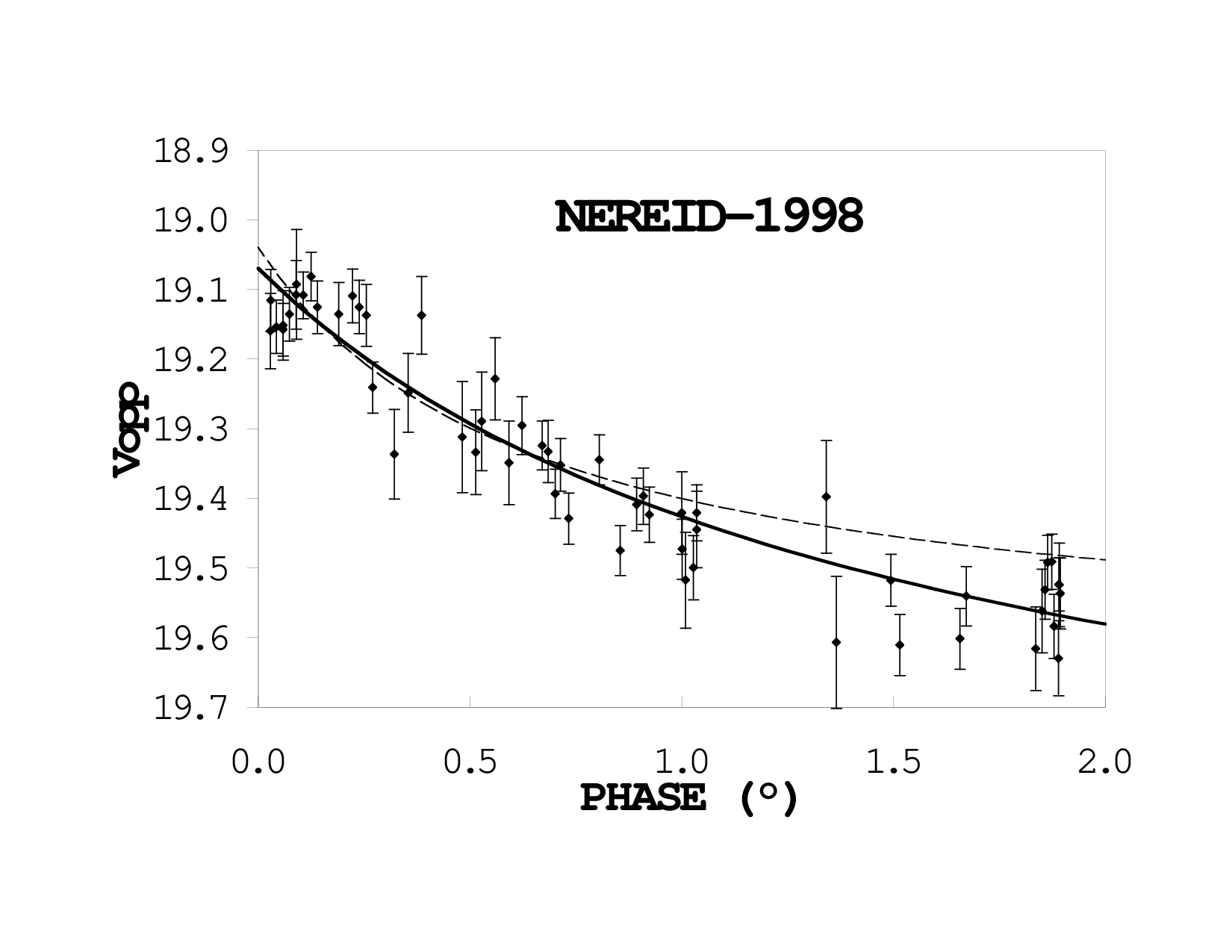}
\vspace{-0.3in}
\plotone{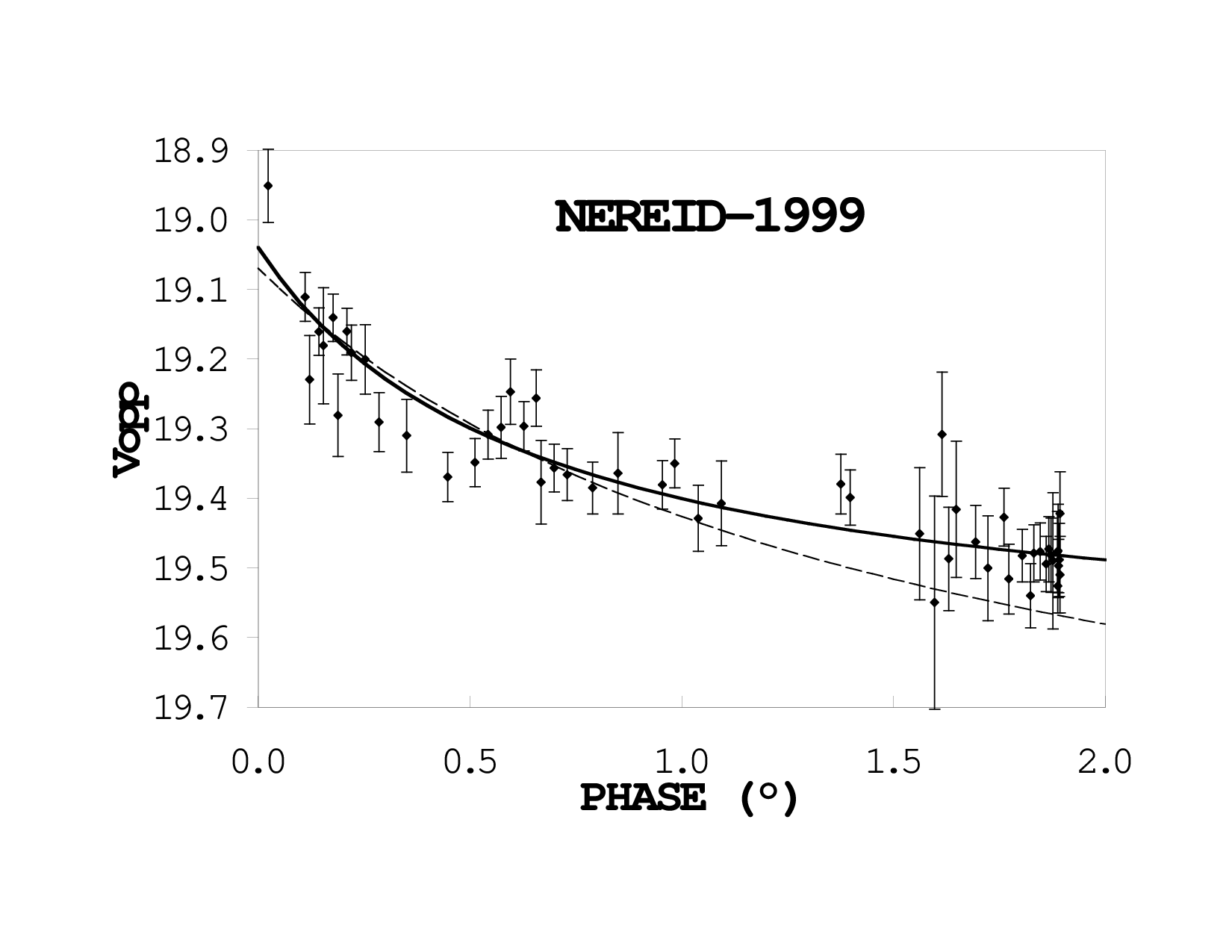}
\vspace{-0.3in}
\plotone{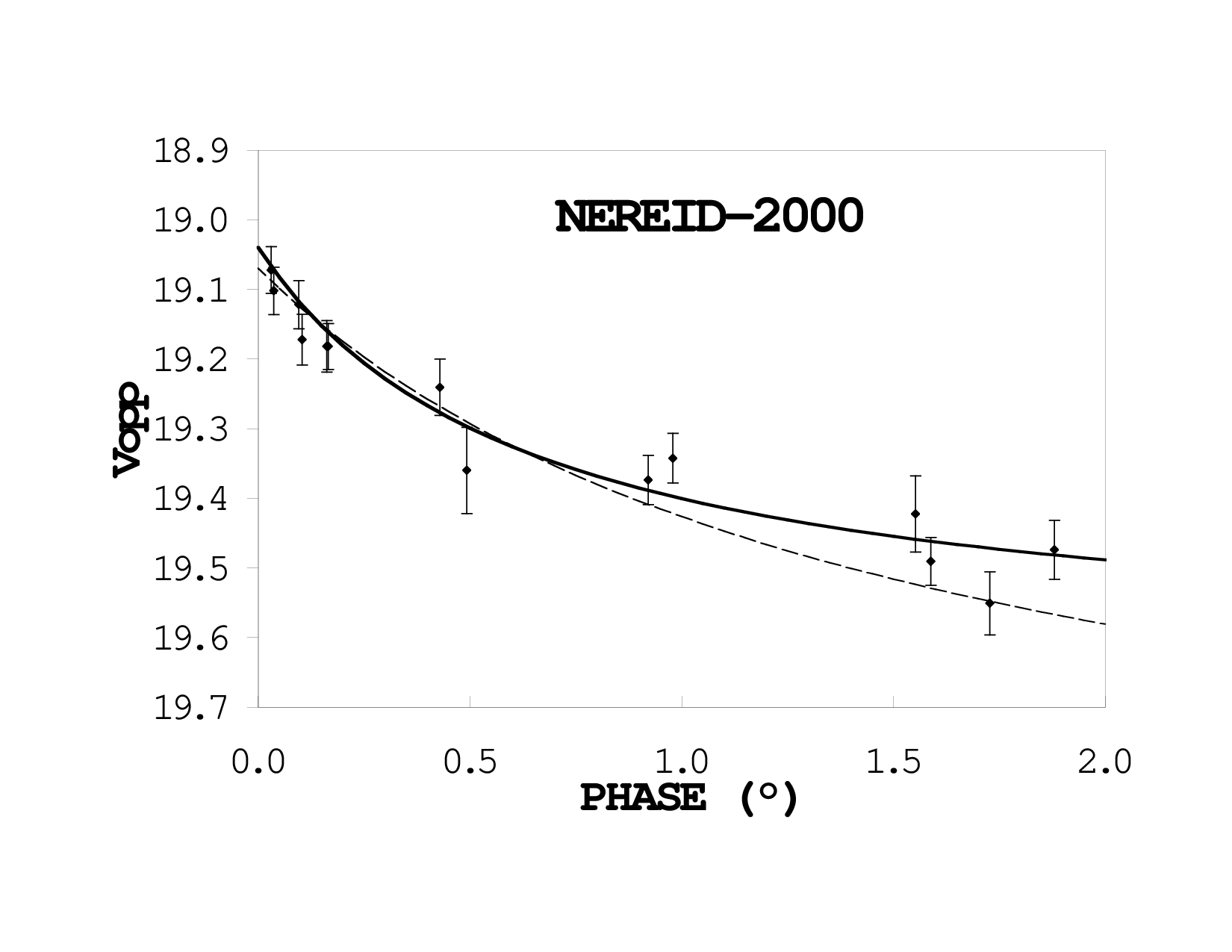}
\vspace{-0.1in}
\columnwidth=2\columnwidth
\caption{
Nereid's phase functions in 1998, 1999, and 2000.
We have well-measured phase functions for Nereid in 1998 (top panel, from 
\citet{sct01}), 1999 (middle panel), and 2000 (bottom 
panel).  Also plotted in each panel are the two Hapke models for the best 
fits to the 1998 and 1999+2000 data (with the bold solid curve being for 
the year of the panel and the light dashed curve for the other fit for 
comparison).  The main point of this figure is that the phase function has 
changed from 1998 to 1999+2000.  For $\alpha < 0.3 \degr$, the 1998 phase 
function is almost flat with a slope of $0.12 \pm 0.12$ mag deg$^{-1}$. 
while the 1999+2000 phase function is very peaked with a slope of 
$0.73 \pm 0.14$ mag deg$^{-1}$.  For $\alpha > 0.85 \degr$, the 1998 
magnitudes consistently and significantly lie below the 1999+2000 model 
while the 1999+2000 magnitudes consistently and significantly lie above 
the 1998 model.  
\label{fig2}}
\end{figure}


\begin{thebibliography}{}

\bibitem[Brown, \& Webster (1998)]{brw98}Brown, M. J. I., \& R. L. Webster 
1998.  A search for distant satellites of Neptune.  Publ. Astron. Soc. 
Australia 15, 326-327.
\bibitem[Buratti, Gibson, \& Mosher (1992)]{bgm92} Buratti, B. J., J. 
Gibson, \& J.A. Mosher 1992. CCD Photometry of the Uranian satellites. \aj 
~104, 1618-1622.
\bibitem[Buratti, Goguen, \& Mosher (1997)]{bgm97} Buratti, B. J., J. D. 
Goguen, \& J. A. Mosher 1997.  No large brightness variations on Nereid.  
Icarus 126, 225-228.
\bibitem[Bus, \& Larson (1989)]{bul89}Bus, E. S. \& S. Larson 1989.  CCD 
photometry of Nereid.  BAAS 21, 982.
\bibitem[Dobrovolskis (1995)]{dob95}Dobrovolskis, A. 1995, Chaotic 
rotation of Nereid?  Icarus 118, 181-198.
\bibitem[Doningue et al. (1991)]{dom91}Domingue, D. L., B. W. Hapke, G. W. 
Lockwood, \& D. T. Thompson 1991.  Europa's phase curve: Implications for 
surface structure.  Icarus 90, 30-42.
\bibitem[Hapke (1983)]{hap83}Hapke, B. 1983.  The opposition effect.  
BAAS 15, 856-857.
\bibitem[Hapke (1993)]{hap93} Hapke, B. 1993, Theory of reflectance and 
emittance spectroscopy.  Cambridge: Cambridge Univ. Press.
\bibitem[Helfenstein, Veverka, \& Hillier (1997)]{hvh97}Helfenstein, J. 
D., J. Veverka, \& J. Hillier 1997.  The Lunar opposition effect: a test 
of alternative models.  Icarus 128, 2-14.
\bibitem[Helfenstein et al. (1998)]{hel98}Helfenstein, J. D. and 21 
colleagues 1998.  Galileo observations of Europa's opposition effect.  
Icarus 135, 41-63.
\bibitem[Klavetter (1989)]{kla89}Klavetter, J. J. 1989.  Rotation of 
Hyperion.  II. Dynamics.  \aj~98, 1855-1874.
\bibitem[Schaefer \& Schaefer (2000)]{scs00} Schaefer, B. E., \& M. W. 
Schaefer 2000.  Nereid has large-amplitude photometric variability.  
Icarus, 146, 541-555.
\bibitem[Schaefer \& Tourtellotte (2001)]{sct01} Schaefer, B. E., \& S. W. 
Tourtellotte 2001. Photometric Light Curve for Nereid in 1998: A Prominent 
Opposition Surge.  Icarus, 151, 112-117.
\bibitem[Schaefer \& Schaefer (1988)]{scs88}Schaefer, M. W., \& B. E. 
Schaefer 1988. Large-amplitude photometric variations of Nereid.  Nature 
333, 436-438.
\bibitem[Schaefer \& Schaefer (1995)]{scs95} Schaefer, M. W., \& B. E. 
Schaefer 1995.  Nereid: Does its continued unique variability indicate a 
Kuiper belt origin?.  BAAS 27, 1167.
\bibitem[Williams, Jones, \& Taylor (1991)]{wjt91}Williams, I. P., D. H. 
P. Jones, \& D. B. Taylor 1991 \mnras~250, 1-2.


\end{thebibliography}
\end{document}